\documentclass[12pt]{article}
\usepackage{epsf}

\begin{document}

\newcommand{\gtrsim}{\mathop{}_{\textstyle \sim}^{\textstyle >}}
\newcommand{\lesssim}{\mathop{}_{\textstyle \sim}^{\textstyle <} }

\newcommand{\rem}[1]{{\bf #1}}


\newcommand{\rs}        {\sqrt{s}}
\newcommand{\ra}        {\rightarrow}   
\newcommand{\Ecm}       {E_{\mathrm{cm}}}
\newcommand{\Ebeam}     {E_{\mathrm{b}}}
\newcommand{\ee}        {{\mathrm e}^+ {\mathrm e}^-}
\newcommand{\ppbar}     {p \bar{p}}
\newcommand{\pp}        {pp}

\newcommand{\Evis}      {E_{\mathrm{vis}}}
\newcommand{\Rvis}      {E_{\mathrm{vis}}\,/\roots}
\newcommand{\Mvis}      {M_{\mathrm{vis}}}
\newcommand{\Rbal}      {R_{\mathrm{bal}}}
\newcommand{\mjet}      {\bar{M}_{\mathrm{jet}}}
\newcommand{\mET}       {$\not \!\!\! {E_T}$\ }
\newcommand{\pt}        {$p_{T}$}
\newcommand{\degree}    {^\circ}
%
%
\newcommand{\thrust}    {T}
\newcommand{\nthrust}   {\hat{n}_{\mathrm{thrust}}}
\newcommand{\thethr}    {\theta_{\,\mathrm{thrust}}}
\newcommand{\phithr}    {\phi_{\mathrm{thrust}}}
\newcommand{\acosthr}   {|\cos\thethr|}
\newcommand{\thejet}    {\theta_{\,\mathrm{jet}}}
\newcommand{\acosjet}   {|\cos\thejet|}
\newcommand{\thmiss}    { \theta_{\mathrm{miss}} }
\newcommand{\cosmiss}   {| \cos \thmiss |}
\newcommand{\fbinv}     {\mathrm{fb}^{-1}}

\newcommand{\phiacop}       {\phi _{\mathrm {acop}}}
\newcommand{\cosjet}        {\cos\thejet}
\newcommand{\costhr}        {\cos\thethr}

%
%
\newcommand{\ff}            {{\mathrm f} \bar{\mathrm f}}
\newcommand{\allqq}         {\sum_{q \neq t} q \bar{q}}
\newcommand{\qq}            {{\mathrm q}\bar{\mathrm q}}
\newcommand{\bb}            {{\mathrm b}\bar{\mathrm b}}
\newcommand{\toppair}        {t\bar{t}}

\newcommand{\tq}            {\mathrm t} 
\newcommand{\bq}            {\mathrm b} 
\newcommand{\cq}            {\mathrm c} 
\newcommand{\qu}            {\mathrm q} 
\newcommand{\ele}           {\mathrm e} 

\newcommand{\nunu}          {\nu \bar{\nu}}
\newcommand{\mumu}          {\mu^+ \mu^-}
\newcommand{\tautau}        {\tau^+ \tau^-}
\newcommand{\ellell}        {l^+l^-}

\newcommand{\Zboson}        {Z^{0}}
\newcommand{\Zv}            {{\mathrm Z}^{*}}
\newcommand{\Wpm}           {W^{\pm}}
\newcommand{\Wp}            {W^+}
\newcommand{\WW}            {{\mathrm W}^+{\mathrm W}^-}
\newcommand{\ZZ}            {{\mathrm ZZ}}
\newcommand{\gv}            {\gamma^*}
\newcommand{\Zg}            {{\mathrm Z} \gamma}
\newcommand{\Zgv}           {{\mathrm Z} \gamma^*}
\newcommand{\Zvgv}          {{\mathrm Z^*} \gamma^*}

\newcommand{\HSM}           {\mathrm H^{0}_{SM}}
\newcommand{\MHSM}          {\mathrm M(H^{0}_{SM})}

\newcommand{\eetautau}      {\ee-\rightarrow {\tau^+}{\tau^-}}
\newcommand{\nulqq}         {\nu \ell {\mathrm q} \bar{\mathrm q}'}
\newcommand{\Wenu}          {{\mathrm{We}} \nu}


\newcommand{\mt}            {M_{\mathrm t}}
\newcommand{\mb}            {M_{\mathrm b}}
\newcommand{\mc}            {M_{\mathrm c}}
\newcommand{\MZ}            {M_{\mathrm Z}}
\newcommand{\MW}            {M_{\mathrm W}}

%
%
\newcommand{\hsusy}         {h^{0}}
\newcommand{\Hsusy}         {\mathrm H^{0}}
\newcommand{\Asusy}         {\mathrm A^{0}}
\newcommand{\CHsusy}         {\mathrm H^{\pm}}

\newcommand{\sfermion}      {\tilde{\mathrm f}}
\newcommand{\sfpair}        {\tilde{\mathrm{f}}^{+} \bar{\tilde{\mathrm{f}}}^{-}}
\newcommand{\sele}          {\tilde{\mathrm e}}
\newcommand{\smu}           {\tilde{\mu}}
\newcommand{\stau}          {\tilde{\tau}}
\newcommand{\staum}         {\tilde{\tau}_{1}}
\newcommand{\sell}          {\tilde{\ell}}
\newcommand{\slepton}       {\tilde{\ell}^{\pm}}
\newcommand{\sellsell}      {\sell^+ \sell^-}
\newcommand{\snu}           {\tilde{\nu}}

\newcommand{\nt}            {\tilde{\chi}^0}
\newcommand{\neutralino}    {\tilde{\chi }^{0}_{1}}
\newcommand{\neutrala}      {\tilde{\chi }^{0}_{2}}
\newcommand{\neutralb}      {\tilde{\chi }^{0}_{3}}
\newcommand{\neutralc}      {\tilde{\chi }^{0}_{4}}
\newcommand{\bino}          {\tilde{\mathrm B}^{0}}
\newcommand{\wino}          {\tilde{\mathrm W}^{0}}
\newcommand{\higgsino}      {\tilde{\mathrm H}^{0}}
\newcommand{\higginoa}      {\tilde{\mathrm H_{1}}^{0}}
\newcommand{\higginob}      {\tilde{\mathrm H_{1}}^{0}}

\newcommand{\ch}            {\tilde{\chi}^\pm}
\newcommand{\chp}           {\tilde{\chi}_{1}^+}
\newcommand{\chm}           {\tilde{\chi}_{1}^-}
\newcommand{\chpm}          {\tilde{\chi}_{1}^\pm}
\newcommand{\cwino}         {\tilde{\mathrm W}^{\pm}}
\newcommand{\chiggsino}     {\tilde{\mathrm H}^{\pm}}
\newcommand{\chargino}      {\tilde{\chi }^{\pm}_{1}}
\newcommand{\charginop}     {\tilde{\chi }^{+}_{1}}
\newcommand{\gra}           {\tilde{\mathrm G}}
\newcommand{\squark}        {\tilde{\mathrm q}}
\newcommand{\gluino}        {\tilde{g}}
\newcommand{\supq}          {\tilde{\mathrm u}}
\newcommand{\sdownq}        {\tilde{\mathrm d}}
\newcommand{\stopm}         {\tilde{\mathrm{t}}_{1}}
\newcommand{\stops}         {\tilde{\mathrm{t}}_{2}}
\newcommand{\stopbar}       {\bar{\tilde{\mathrm{t}}}_{1}}
\newcommand{\stopx}         {\tilde{\mathrm{t}}}
\newcommand{\stopl}         {\tilde{\mathrm{t}}_{\mathrm L}}
\newcommand{\stopr}         {\tilde{\mathrm{t}}_{\mathrm R}}
\newcommand{\stoppair}      {\tilde{\mathrm{t}}_{1} \bar{\tilde{\mathrm{t}}}_{1}}
\newcommand{\sbotm}         {\tilde{\mathrm{b}}_{1}}
\newcommand{\sbots}         {\tilde{\mathrm{b}}_{2}}
\newcommand{\sbotbar}       {\bar{\tilde{\mathrm{b}}}_{1}}
\newcommand{\sbotx}         {\tilde{\mathrm{b}}}
\newcommand{\sbotl}         {\tilde{\mathrm{b}}_{\mathrm L}}
\newcommand{\sbotr}         {\tilde{\mathrm{b}}_{\mathrm R}}
\newcommand{\sbotpair}      {\tilde{\mathrm{b}}_{1} \bar{\tilde{\mathrm{b}}}_{1}}
%
%
\newcommand{\MGUT}          {M_{\mathrm{GUT}}}
\newcommand{\mscalar}       {m_{0}}
\newcommand{\Mgaugino}      {M_{1/2}}
\newcommand{\tanb}          {\tan \beta}

\newcommand{\mixstop}       {\theta _{\stopx}}
\newcommand{\mixsbot}       {\theta _{\sbotx}}
\newcommand{\mchar}         {m_{\chpm}}
\newcommand{\mstop}         {m_{\stopm}}
\newcommand{\msbot}         {m_{\sbotm}}
\newcommand{\mchi}          {m_{\neutralino}}

\renewcommand{\thefootnote}{\fnsymbol{footnote}}
\setcounter{footnote}{0}
\begin{titlepage}

\def\thefootnote{\fnsymbol{footnote}}

\begin{center}

\hfill UT-ICEPP-07-02\\
\hfill TU-792\\
\hfill May, 2007\\

\vskip .75in

{\Large \bf 
Testing the Anomaly Mediation at the LHC
}

\vskip .75in

{\large
$^{(a)}$Shoji Asai,
$^{(b)}$Takeo Moroi, 
$^{(a)}$Kazuyuki Nishihara, 
and $^{(a)}$T.T. Yanagida
}

\vskip 0.25in

{\em $^{(a)}$Department of Physics, University of Tokyo,
Tokyo 113-0033, Japan
}

\vskip 0.2in

{\em $^{(b)}$Department of Physics, Tohoku University,
Sendai 980-8578, Japan
}

\end{center}
\vskip .5in

\begin{abstract}

We consider a supersymmetric model in which gaugino masses are
generated by the anomaly-mediation mechanism while scalar masses are
from tree-level supergravity interaction.  In such a model, scalar
fermions as well as Higgsinos become as heavy as 
$O(10-100\ {\rm TeV})$ and hence only the
gauginos are 
superparticles kinematically accessible to the LHC.  
We study how and
how well the properties of gauginos can be studied.  We also discuss
the strategy to test the anomaly-mediation model at the LHC.

\end{abstract}

\end{titlepage}

\renewcommand{\thepage}{\arabic{page}}
\setcounter{page}{1}
\renewcommand{\thefootnote}{\#\arabic{footnote}}
\setcounter{footnote}{0}

\section{Introduction}

Anomaly mediation of supersymmetry (SUSY) breaking in a hidden to the
SUSY standard-model (SUSY SM, or SSM) sector
\cite{Giudice:1998xp,Randall:1998uk} is very attractive, since it is
the simplest mechanism for the mediation of SUSY breaking. Namely, the
anomaly mediation always takes place in a generic SUSY theory and
hence we do not need any extra assumption beside the presence of SUSY
breaking sector to mediate the SUSY-breaking effects to the SSM sector.
Without any special requirement on the K\"ahler potential, but just by
assuming that there is no singlet field in the SUSY breaking sector,
the scenario of anomaly mediation predicts the so-called split SUSY
spectrum \cite{splitSUSY} where squarks and sleptons may have masses
of the order 10 $-$ 100 TeV 
while the masses of gauginos are in the range of
100 GeV $-$ 1 TeV \cite{Wells:2004di,Ibe:2006de}.\footnote
{In this letter, we use ``anomaly-mediation model'' for those where
  gaugino masses are generated by the effect of anomaly mediation while
  the scalar masses are from tree-level supergravity interaction between
  observable-sector fields and SUSY breaking fields.}

Because of the relatively large masses of squarks we need a very
precious fine-tuning of parameters to obtain the correct electroweak
symmetry breaking, but on the other hand it solves many serious
problems in the SSM. First of all the flavor-changing neutral current
and CP-violation problems become very milder due to the large masses
of squarks and sleptons.  We may naturally explain no discovery of
Higgs at LEP and no discovery of proton decays induced by
dimension-five operators. Furthermore, the gravitino mass is also
predicted at the order of 100 TeV, which makes the cosmological
gravitino problems much less severe \cite{gravitino100TeV, KawTakYan}.
In fact, it has been pointed out that the leptogenesis
\cite{Leptogenesis} does work in the anomaly-mediation model, since
the reheating temperature $T_R$ can be as high as $10^{10}$ GeV
without any conflict with cosmology \cite{Ibe:2004tg}.

Although the squarks and sleptons are so heavy, the masses of gauginos
may be in the accessible range to the LHC experiments.  Thus, even in
this model, discovery of the signals from the productions of
superparticles may be possible at the LHC.  More importantly, the
anomaly-mediation model predicts unique mass relation among gauginos, as
we will see in the following.  In particular, the anomaly mediation
predicts $m_{\tilde{W}}<m_{\tilde{B}}<m_{\tilde{g}}$ (with
$m_{\tilde{B}}$, $m_{\tilde{W}}$, and $m_{\tilde{g}}$ being the gaugino
masses of $U(1)_Y$, $SU(2)_L$, and $SU(3)_C$ gauge groups, respectively)
in a large region of the parameter space \cite{Ibe:2006de}.  
In this letter, we discuss how the anomaly-mediation model can be
studied at the LHC.

\section{Model}
\label{sec:model}

Let us start our discussion by summarizing basic features of the
model.  As we have mentioned in the previous section, we consider the
anomaly-mediation model in which all the sfermions as well as heavy
Higgses and Higgsinos are heavy (of masses of the order of
$100\ {\rm TeV}$).  Assuming that there is no singlet field in the
SUSY breaking sector, all the gaugino masses are suppressed compared
to the SUSY breaking scalar masses.

In this class of models, gaugino masses are mainly from the effect of
anomaly mediation and possible radiative correction due to the
Higgs-Higgsino loop \cite{Giudice:1998xp,Gherghetta:1999sw}.  Then,
the SUSY breaking gaugino mass parameters (at the scale of sfermion
masses $m_{\tilde{f}}$) are obtained as
\begin{eqnarray}
 M_1 &=& \frac{g_1^{2}}{16 \pi^{2}} \left( 11 m_{3/2} + L \right),
  \label{M1}
  \\
 M_2 &=& \frac{g_2^{2}}{16 \pi^{2}} \left( m_{3/2} + L \right),
  \\
 M_3 &=& \frac{g_3^{2}}{16 \pi^{2}} \left( -3 m_{3/2} \right),
  \label{M3}
\end{eqnarray}
where $g_1$, $g_2$, and $g_3$ are gauge coupling constants of
$U(1)_Y$, $SU(2)_L$, and $SU(3)_C$ gauge groups, respectively, and
$m_{3/2}$ is the gravitino mass.  (Hereafter, we use the convention
such that $m_{3/2}$ is real and positive.)  In addition,
\begin{eqnarray}
 L \equiv \mu \sin2\beta 
  \frac{m_{A}^{2}}{|\mu|^{2}-m_{A}^{2}} \ln \frac{|\mu|^{2}}{m_{A}^{2}},
  \label{L-parameter}
\end{eqnarray}
with $\tan\beta$ being the ratio of two Higgs bosons, and $m_A$ the
mass of heavy Higgses.  Thus, due to the Higgs-Higgsino loop
contribution, gaugino masses may significantly deviate from the
pure-anomaly-mediation relation (which is given by taking $L=0$).
However, since the natural sizes of $|\mu|$ and $m_A$ are both of the
order of the gravitino mass, the $L$-parameter is expected to be at
most $O(m_{3/2})$.  (Notice that $|L|\leq m_A$.)  Consequently, with a
natural choice of $\mu$- and $B_\mu$-parameters, Wino becomes the
lightest among gauginos in a large region of the parameter space.  In
this case, taking account of the radiative correction due to
electroweak gauge bosons, neutral Wino $\tilde{W}^0$ becomes the
lightest superparticle (LSP) \cite{Feng:1999fu}.  Charged Wino
$\tilde{W}^\pm$ becomes slightly heavier than the neutral one; the
mass splitting is as small as $m_{\tilde{W}^\pm}-m_{\tilde{W}^0}\simeq
155-170\ {\rm MeV}$.  As we will see, the smallness of the mass
splitting has an important implication to the study of the
anomaly-mediation model at the LHC.

We should also note that three gaugino masses depend on three
parameters: $m_{3/2}$, $|L|$, and ${\rm Arg}(L)$.  Even so, non-trivial
constraint exists among the gaugino masses.  In order to see how the
gaugino masses are constrained, it is instructive to approximate the
physical (i.e., on-shell) gaugino masses by $M_1$ $-$ $M_3$ given in
Eqs.\ (\ref{M1}) $-$ (\ref{M3}).  Then, we can see
\begin{eqnarray}
  \left| \frac{10 g_1^{2}}{3g_3^{2}} m_{\tilde{g}}
    - \frac{g_1^{2}}{g_2^{2}} m_{\tilde{W}} \right|
  \lesssim m_{\tilde{B}} \lesssim
  \frac{10 g_1^{2}}{3g_3^{2}} m_{\tilde{g}}
    + \frac{g_1^{2}}{g_2^{2}} m_{\tilde{W}}.
\end{eqnarray}
Thus, once the Wino and gluino masses are fixed, upper and lower bounds
on the Bino mass are obtained.  (In our quantitative analysis in the
following sections, we calculate the bounds more accurately by taking
into account the renormalization-group effect below the sfermion-mass
scale $m_{\tilde{f}}$, taking $m_{\tilde{f}}=m_{3/2}$.  We have checked
that the dependence on $m_{\tilde{f}}$ is rather mild; the bounds change
$\sim 10\ {\rm GeV}$ when $m_{\tilde{f}}$ is varied in the range
$m_{3/2}/2<m_{\tilde{f}}<2m_{3/2}$.)  It is an important test of the
anomaly-mediation model to see if the gauginos satisfy the mass
relation.

\section{Anomaly-Mediation Model at the LHC}

\subsection{Set up}

Because the gauginos are the only superparticles kinematically
accessible to the LHC, and also because an important information is
imprinted in the gaugino masses, we must study the properties of the
gauginos at the LHC if the anomaly-mediation model is realized in
nature.  In the following, we discuss how well the gauginos can be
investigated at the LHC.  As an example, in this letter, we consider the
case where the underlying model gives $m_{\tilde{B}}=400\ {\rm GeV}$,
$m_{\tilde{W}}=200\ {\rm GeV}$, and $m_{\tilde{g}}=1\ {\rm TeV}$ (which
is given by $m_{3/2}\simeq 39\ {\rm TeV}$, $|L|\simeq 28\ {\rm TeV}$,
and ${\rm Arg}(L)=0$).  Notice that the neutral Wino is the LSP.

First, we comment on the decay of the charged Wino.  Since the mass
splitting $m_{\tilde{W}^\pm}-m_{\tilde{W}^0}$ is very small,
$\tilde{W}^\pm$ decays into $\tilde{W}^0$ emitting an extremely soft
pion.  Thus, it is not easy to identify the decay of the charged Wino at
the LHC.  In addition, since the typical decay length of
$\tilde{W}^\pm$ is $O(1-10\ {\rm cm})$, it is highly challenging to find
the track of the charged Wino although it may not be impossible.  Thus,
in our main discussion, we make a conservative assumption that we cannot
find the charged-Wino track; then both $\tilde{W}^0$ and $\tilde{W}^\pm$
are treated as invisible particles.  However, we will also discuss the
implications of discovering the charged-Wino tracks.

\begin{figure}
    \centerline{\epsfxsize=0.5\textwidth\epsfbox{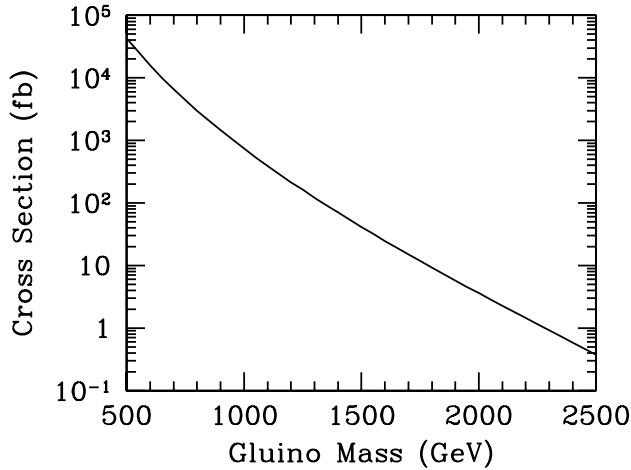}}
    \caption{Cross section for the process $pp\rightarrow
      \tilde{g}\tilde{g}$ for $\sqrt{s}=14\ {\rm TeV}$.  Here the
      squarks are assumed to be heavy enough to be neglected.}
  \label{fig:pp2gogo}
\end{figure}

The most important production process of superparticles at the LHC is
the pair production of gluinos $\tilde{g}$'s: $pp\rightarrow
\tilde{g}\tilde{g}$, as far as the gluino mass is about $1\ {\rm TeV}$
or smaller; in Fig.\ \ref{fig:pp2gogo} we plot the cross section for
this process.  When $m_{\tilde{g}}\simeq 1\ {\rm TeV}$, for
example, thousands of supersymmetric events will be available at the
LHC with ${\cal L}=10\ {\rm fb^{-1}}$ (with ${\cal L}$ being
luminosity).

Once gluino is produced, it decays into a lighter gaugino (i.e.,
Wino $\tilde{W}$ or Bino $\tilde{B}$) and standard-model fermions.  If
Bino $\tilde{B}$ is produced by the decay of gluino, it decays
successively to Wino.  Experimental signals strongly depend on how the
gauginos decay.  The decay patterns of gauginos are sensitive to the
masses of sfermions and Higgsinos, and hence are not predictable.  Thus,
we make several assumptions for our study.  For the decay of gluino, we
adopt
\begin{eqnarray}
Br(\tilde{g}\rightarrow \tilde{B}q\bar{q})
=1-Br(\tilde{g}\rightarrow \tilde{W}q\bar{q})=0.25,
\end{eqnarray}
where $q$ denotes the standard-model quarks.

On the decay of Bino, important processes are $\tilde{B}\rightarrow
\tilde{W}^\pm W^\mp$ via Bino-Wino mixing, 
$\tilde{B}\rightarrow\tilde{W}^0 h_{\rm SM}$ (with $h_{SM}$ being
standard-model-like Higgs boson)
via Bino-Higgsino mixing,
and $\tilde{B}\rightarrow\tilde{W}f\bar{f}$ 
(with $f$ being standard-model fermions) due to
sfermion-exchanges.  In a large fraction of the parameter space,
$\tilde{B}\rightarrow \tilde{W}^\pm W^\mp$ and
$\tilde{B}\rightarrow\tilde{W}^0 h_{\rm SM}$
become the dominant decay
modes.\footnote
{We thank T. Watari for a useful discussion on the Bino decay.}
If so, however, it is difficult to study the properties of $\tilde{B}$.
Then, probably the anomaly-mediation model can be tested only by finding
the charged-Wino track, which will be challenging, (as well as by the
measurement of $m_{\tilde{g}}-m_{\tilde{W}}$, which we discuss below).
If there exists a significant hierarchy between sfermion and Higgsino
masses, however, the last process may dominate.  In particular, if the
sleptons are much lighter than Higgsino (by factor $10$ or so),
$\tilde{B}\rightarrow \tilde{W}l^+l^-$ may acquire a large branching
ratio.  As we will see, invariant-mass distribution of $l^+l^-$ gives an
important information on the Bino mass.  Thus, in order to demonstrate
what we can learn from the decay mode $\tilde{B}\rightarrow
\tilde{W}^0l^+l^-$, we consider a special case where sleptons are much
lighter than the Higgsino so that the decay modes, $\tilde{B}\rightarrow
\tilde{W}^\pm W^\mp$ and 
$\tilde{B}\rightarrow\tilde{W}^0 h_{\rm SM}$,
become negligible.  For our simulation, we use
\begin{eqnarray}
Br(\tilde{B}\rightarrow \tilde{W}L\bar{L})
=1-Br(\tilde{B}\rightarrow \tilde{W}q\bar{q})=0.3,
\end{eqnarray}
where $L=\nu,l^-$ is for the standard-model leptons.

\subsection{Simulated background samples and selections}

The gluino pair is produced mainly with gluon-gluon collision, and the
gluino decays into $\tilde{W}$ or $\tilde{B}$ with two or four high \pt
\, jets.  Therefore the missing transverse energy, \mET, carried away by
two Winos plus four high \pt\ jets is the leading experimental signature
(defined as ``no lepton mode'').  Also lepton pair from the decays of
$\tilde{B}$, four-jet and \mET is the next leading signature of the
signal (defined as ``dilepton mode'').  Fraction of the event with
dilepton depends strongly on the decay branching fraction,
$Br(\tilde{g}\rightarrow \tilde{B}q\bar{q})\times
Br(\tilde{B}\rightarrow \tilde{W}^0{\ellell})$.  The discovery potential
will be mainly determined with the no lepton mode and the dilepton mode
will be used to measure the mass difference between $\tilde{W}$ and
$\tilde{B}$.

The following four standard-model processes can potentially 
have the event topology of \, \mET \, with jets:
\begin{itemize}
\item $\Wpm$ + jets, ($\Wpm \ra \ell \nu$)
\item $\Zboson$ + jets, ($\Zboson \ra \nunu, \tautau $)
\item $t\bar{t}$
\item QCD jets: Heavy flavor quarks ($b$ and $c$) with semi-leptonic
decay and the light flavor jets with mis-measurement.
\end{itemize}

The high \pt\ multi-jets are key of the analysis and 
they should be estimated with the Matrix-Element calculation. 
Parton-Shower is not good approximation in such a high \pt\ region
and the background contributions would be underestimated~\cite{asai}\@.
These background processes are generated with ALPGEN2.05~\cite{ALPGEN}, 
and the exact Matrix-Element Calculations are applied up-to 
five partons. 
The produced 200 M events are fed into the Parton-Shower 
generator (JIMMY4.0/Herwig6.5~\cite{Herwig})
in order to evolute the QCD shower.
Multiple-interaction processes are also taken into account. 
Matching between Matrix-Element and the evaluated 
Parton-Shower are also applied to remove the double counts.
The detector effect is taken into account using the smearing 
Monte Carlo simulation of the ATLAS detector (ATLFAST~\cite{TDR1}).

\begin{figure}
    \centerline{\epsfxsize=0.5\textwidth\epsfbox{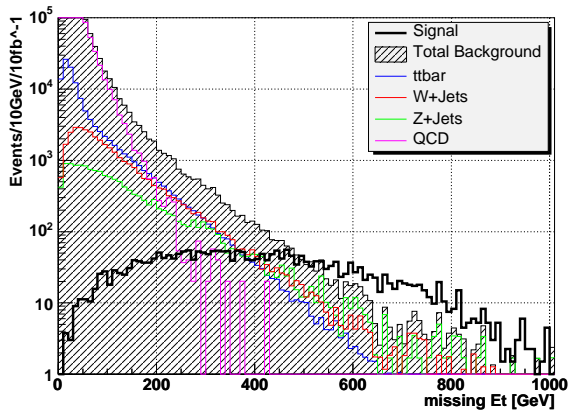}}
    \caption{\mET distribution of the SUSY signal and background
             processes: The open histogram shows the SUSY signal (with
             $m_{\tilde{g}}=1\ {\rm TeV}$) and the hatched shows the
             sum of background distributions.}
  \label{fig:met}
\end{figure}

The following simple event selections are applied, 
\begin{itemize}
\item Number of jets with \pt $>$ 50~GeV is larger than or equal to 4.
\item \pt\ of the jets are required to be larger than 200 and 100~GeV
      for the leading and 2nd leading jets, respectively.
\item The missing transverse energy, \mET, is larger than 300~GeV.
Fig.\ \ref{fig:met} shows the \, \mET distributions for the no lepton
mode.
\item The same flavor two leptons with \pt $>$ 20~GeV 
are required for the dilepton mode.
\end{itemize}

\begin{figure}
    \centerline{\epsfxsize=0.4\textwidth\epsfbox{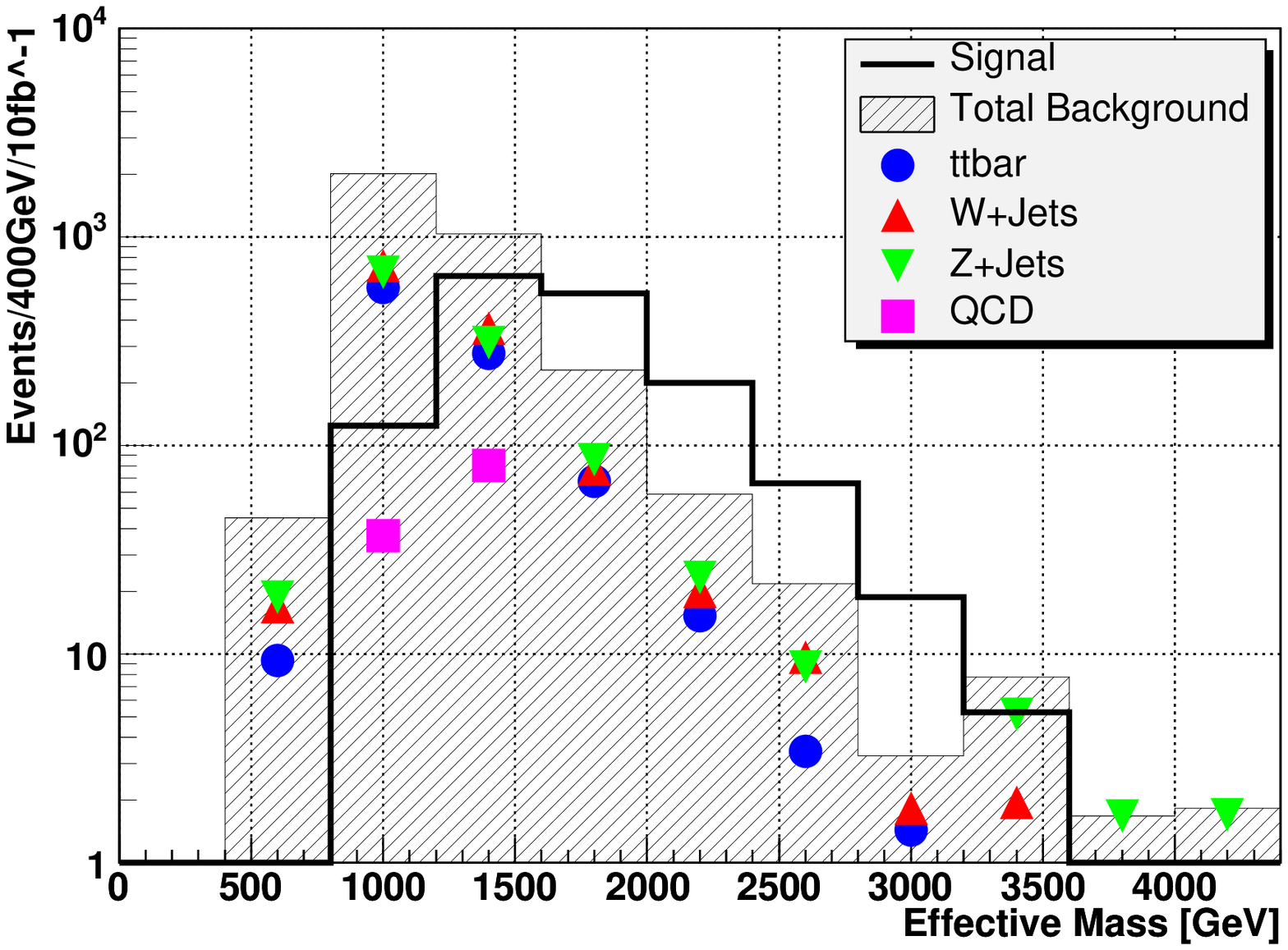}
                \epsfxsize=0.4\textwidth\epsfbox{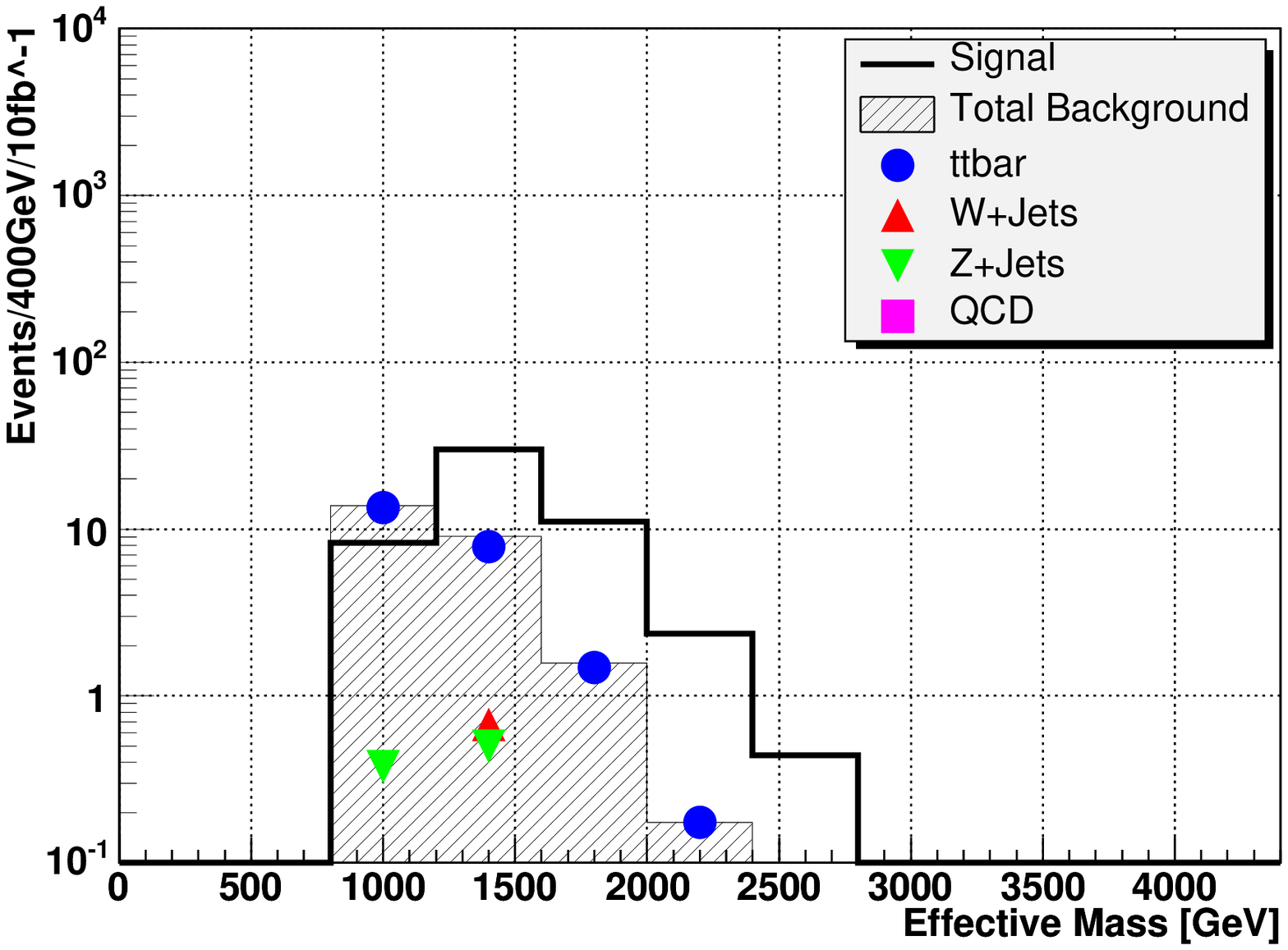}}
    \caption{Effective mass distributions of the SUSY signal and
             background processes: No lepton mode and dilepton mode.
             In both figures, the open histogram shows the SUSY signal
             and the hatched shows the sum of background
             distributions. (Blue circle, red triangle, green triangle
             and magenta box show the top, $W^\pm$ ,$Z^0$ and QCD
             processes, respectively.)}
  \label{fig:meff1}
\end{figure}

The effective mass, which is define as \, \mET $+ \sum_{4jets}$ \pt ,
is a good variable to discriminate the SUSY signal from
the SM background processes, and Figs.\ \ref{fig:meff1}
show the effective mass distributions of the signal mentioned above
and the SM background processes for the no lepton and dilepton modes.
Three standard model processes, $\Wpm$ + jets, $\Zboson$ + jets, and
$\toppair$ contribute equally to no lepton mode, but the signal
excesses (with $m_{\tilde{g}}=1\ {\rm TeV}$) will be observed in the high
effective mass region.  ATLAS has a discovery potential up-to 1.2~TeV
with an integrated luminosity of 10~$\fbinv$.  
$\toppair$ process is the dominant background for two lepton mode.
Since flavor of the selected two leptons are independent in this case,
these background processes can be removed easily using the flavor
subtraction~\cite{ATLASTDRFS}.

\subsection{Mass reconstruction}

Now, let us discuss how and how well the gaugino masses can be
reconstructed.  With the gluino pair production, we obtain, at the
parton level, (i) four quark jets, and (ii) several charged leptons (or
tau jets).  The momentum distribution of the quark jets contains an
information on $m_{\tilde{W}}$ and $m_{\tilde{g}}$, while that of
charged leptons contains that on $m_{\tilde{B}}$ and $m_{\tilde{W}}$.
Indeed, if we concentrate on the decay of single gluino:
$\tilde{g}\rightarrow \tilde{W}q\bar{q}$, invariant mass of the
$q\bar{q}$ system $M_{q\bar{q}}$ is required to be
\begin{eqnarray}
M_{q\bar{q}} \leq m_{\tilde{g}} - m_{\tilde{W}},
\end{eqnarray}
while, for the case of $\tilde{g}\rightarrow \tilde{B}q\bar{q}$,
followed by $\tilde{B}\rightarrow \tilde{W^0}l^+l^-$, we obtain
\begin{eqnarray}
M_{l^+l^-} \leq m_{\tilde{B}} - m_{\tilde{W}},~~~
M_{l^+l^-q\bar{q}} \leq m_{\tilde{g}} - m_{\tilde{W}},
\end{eqnarray}
where $M_{l^+l^-}$ and $M_{l^+l^-q\bar{q}}$ are invariant masses of
$q\bar{q}$ and $l^+l^-q\bar{q}$ systems, respectively.  Thus, from the
invariant-mass distributions, we can extract the information on the
gaugino masses.

\begin{figure}
    \centerline{\epsfxsize=0.5\textwidth\epsfbox{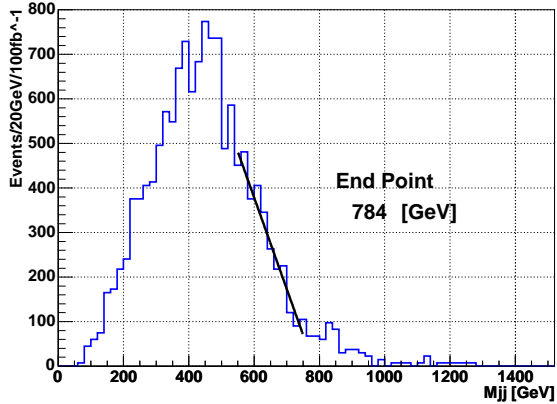}}
    \caption{$M_{jj}$ distribution for the no lepton mode.}
  \label{fig:mqq}
\end{figure}

The same selections mentioned in the previous subsection are applied to
make the mass distributions for $M_{q\bar{q}}$, $M_{l^+l^-}$ and
$M_{l^+l^- q\bar{q}}$. 
The integrated luminosity of 100~$\fbinv$ is
assumed to measure these invariant masses.  The leading four jets are
used to make combination of dijets, and there are three possible
combinations of four jets, (1+3, 2+4), (1+4, 2+3) and (1+2, 3+4).  For
example, (1+3, 2+4) means that the leading \pt\ and 3rd leading jets
are considered to be emitted form the same gluino, and 2nd and 4th leading
\pt\ jets have the same parent.  The fractions of the correct
combination are the same in the combinations of (1+3, 2+4) and (1+4,
2+3).  We select the combination in which the difference between two
calculated invariant masses of dijets is smaller than the other
combination.  On the other hand, the fraction of the wrong combination
is larger in the (1+2, 3+4) combination, so this combination is used
for comparison with the other combinations.  The difference between
two calculated invariant masses is required to be smaller than 100~GeV
in oder to suppress the contamination of $\tilde{g}\rightarrow
\tilde{B}q\bar{q} \rightarrow q\bar{q} q\bar{q}\tilde{W}$.
Fig.\ \ref{fig:mqq} shows the distribution of the invariant mass of
dijets $M_{jj}$, in which the fitted endpoint using linear function is
784$\pm$37~GeV (the expected endpoint is 800~GeV).  Statistical error of
the fitted endpoint is 5\% with an integrated luminosity of
100~$\fbinv$.  Jet is reconstructed with the cone algorithm (size of
0.4), and some part of the evaluated shower escapes from the cone.
This is the reason why the reconstructed $M_{jj}$ becomes smaller than
parton level ($M_{q\bar{q}}$). This shift can be collected with
parton-jet calibrations and the uncertainty in this calibration is
finally 1\%.\footnote
{The calibration of jet energy to parton level will be performed using
$W \ra q\bar{q}$ and $gq\rightarrow\gamma q$ (with $g$ and $\gamma$
being gluon and photon, respectively) events, and finally 1\% level
accuracy will be obtained at ATLAS detector.}
There is also another systematic uncertainty in the fitting procedure.
The other samples with the different mass combinations are also generated, and 
the same selections and fitting procedure are applied. 
The fitted endpoints are proportional to the mass difference at parton level
and consistent within an error of 5\%.

\begin{figure}
 \centerline{\epsfxsize=0.4\textwidth
 \epsfbox{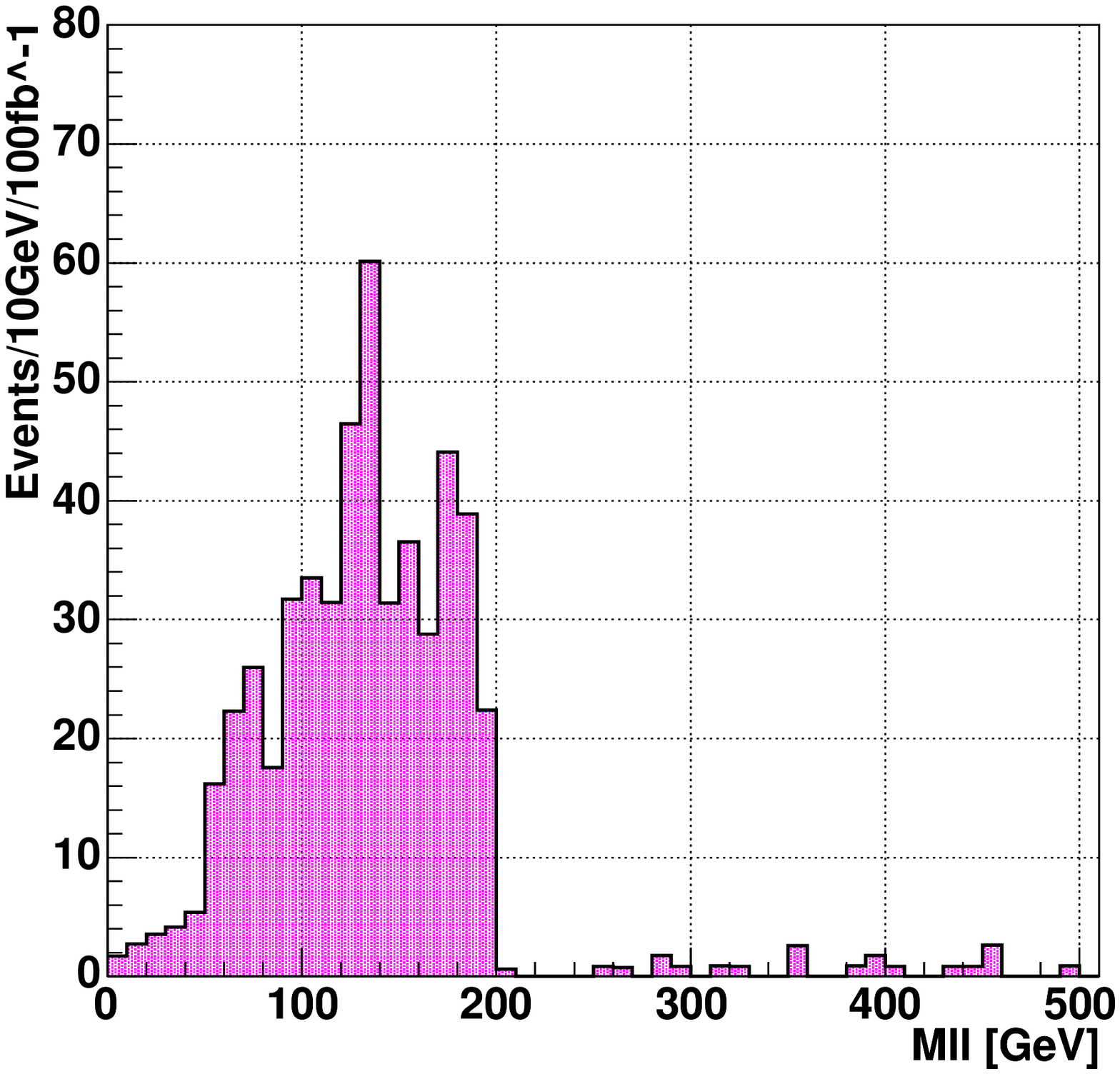}
 \epsfxsize=0.4\textwidth\epsfbox{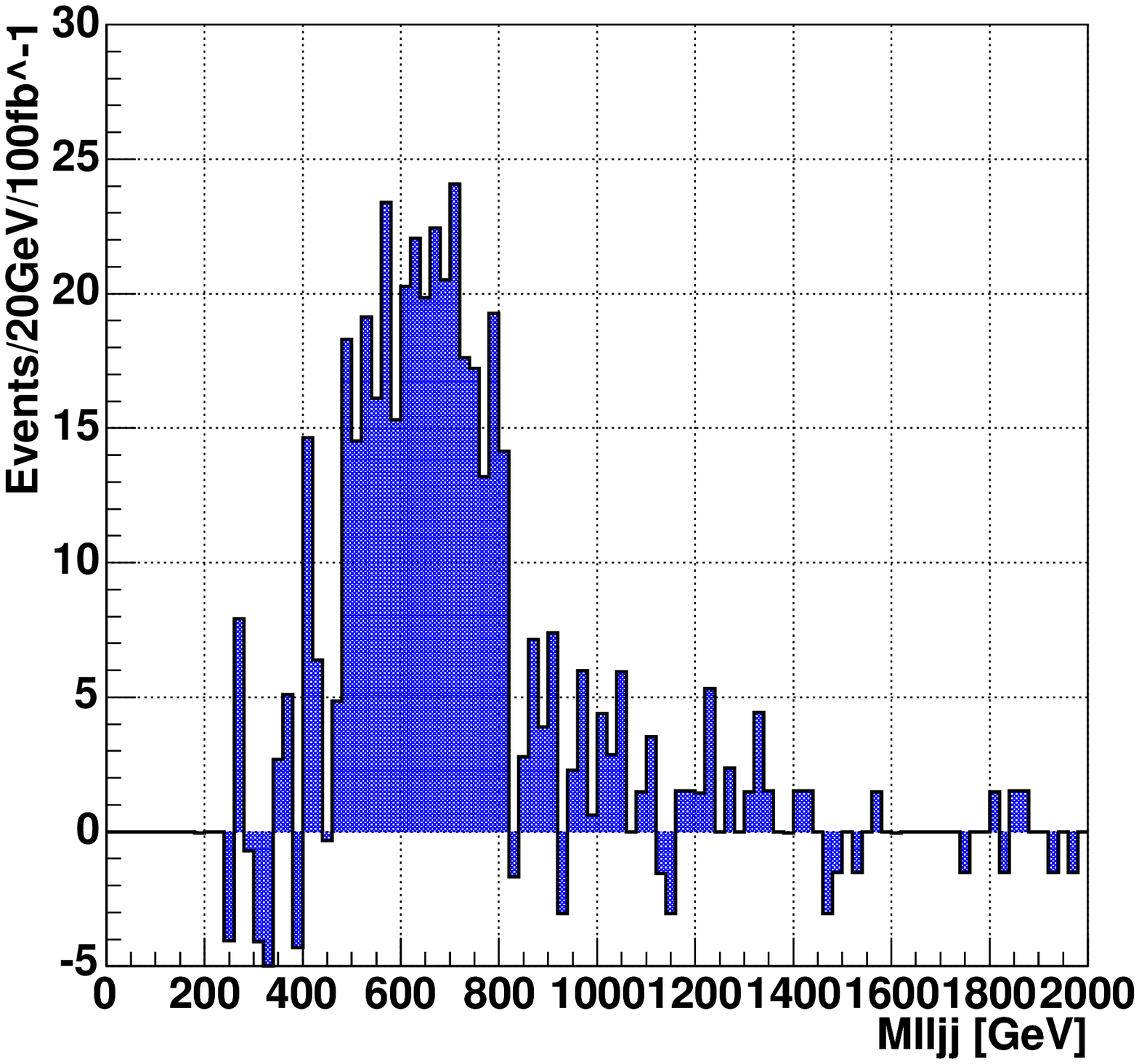}}
 \caption{$M_{\ellell}$ and $M_{\ellell \qq}$ distributions for the 
 dilepton mode. }
  \label{fig:mll}
\end{figure}

Fig.\ \ref{fig:mll}(a) shows the $M_{\ellell}$ distribution for
the dilepton mode after the flavor subtraction~\cite{ATLASTDRFS} is
performed.  Clear edge is observed at the expected
endpoint (200~GeV), and the statistical error of the fitted edge is
1\ \% with an integrated luminosity of 100~$\fbinv$.  Systematic
uncertainty in the absolute energy calibration is much less than the
statistical error.

In order to calculate the invariant mass of dijets and dileptons
the corresponding two jets are selected as follows: 
One of the leading or 2nd leading \pt\ jet is selected,  
which is closer to the dilepton system.
The distance between the dilepton and this jet is also required to be smaller
than 2.0.  The 4th leading jet is considered as another jet, since the
\pt\ becomes smaller due to the smaller mass difference of $\tilde{g}$
and $\tilde{B}$.  The invariant mass between dilepton and the dijets
are shown in Fig.\ \ref{fig:mll}(b), after the flavor subtraction is
applied.  
Clear edge is still observed at 800~GeV, 
and the statistical error is 50~GeV.

\subsection{Other possible information}

So far, we have seen that some of the mass differences can be
determined by the LHC.  In particular, if $Br(\tilde{g}\rightarrow
\tilde{B}q\bar{q})$ and $Br(\tilde{B}\rightarrow \tilde{W}^0{\ellell})$
are both sizable, two mass differences can be reconstructed.  Even so,
the confirmation of the anomaly-mediation mass spectrum of gauginos is
not straightforward because the following questions, which are crucial
to test the anomaly-mediation model, are not answered yet.

First, we have seen that the information on mass differences is
obtained from the invariant-mass distribution of jets and leptons.
However, it is still an open question how the masses of gauginos can
be determined.  In addition, one should note that the event shape from
the decay chain $\tilde{g}\rightarrow\tilde{B}q\bar{q}$ followed by
$\tilde{B}\rightarrow\tilde{W}^0l^+l^-$ (or
$\tilde{B}\rightarrow\tilde{W}^\pm l^\mp\nu$) is hardly distinguished
from that from $\tilde{g}\rightarrow\tilde{W}^0 q\bar{q}$ followed by
$\tilde{W}^0\rightarrow\tilde{B}l^+l^-$ (or
$\tilde{g}\rightarrow\tilde{W}^\pm q\bar{q}'$ followed by
$\tilde{W}^\pm\rightarrow\tilde{B}l^\pm\nu$), in particular 
if the track of the charged Wino is not observable.  Thus, it may not
be easy to confirm that Wino is the LSP.  For the test
of anomaly-mediation model, it is important to answer the following
questions:
\begin{itemize}
\item[1.] How do we determine the masses of gauginos rather than mass
  differences?
\item[2.] How do we know that the lightest state is Wino, not Bino?
\end{itemize}

These can be easily answered if the (short) tracks of charged Wino can
be identified at the LHC \cite{Ibe:2006de}.  Although it is very
challenging, such tracks may be seen in the inner detector, like the
semiconductor pixel detector and/or the transition radiation tracker
(TRT).  If the tracks of charged Winos can be identified, it is obvious
that the lightest gaugino is Wino, not Bino.  In addition, Wino mass may
be determined by using timing information combined with momentum
information on the charged-Wino track.  The resolution of the determined
$\beta$ is about 0.1 if $\beta$ is less than 0.85.  So the mass can be
determined with accuracy of 10\%, if we have enough samples of this
exotic track.

Even if the information on the charged-Wino track is not
available, the above questions may still be answered.  For the first
question, one way to determine the gaugino masses is to use the
cross-section information on the process
$pp\rightarrow\tilde{g}\tilde{g}$.  As one can see in Fig.\
\ref{fig:pp2gogo}, $\sigma(pp\rightarrow\tilde{g}\tilde{g})$ strongly
depends on the gluino mass.  Thus, even if the error in the cross
section is fairly large, we may still have a useful constraint on the
gluino mass.  When the signal number can be determined very roughly
with an accuracy of 20\ \% level, which is the uncertainty of the
background, gluino mass can be estimated with an accuracy of 3\ \%.

In order to discriminate the possibility of Bino-LSP, we may be able
to use the Drell-Yan process $pp\rightarrow\tilde{W}\tilde{W}$.  If
the $\tilde{W}^0$ is the lightest neutralino, such a process does not
provide any signal.  If Bino is the LSP, on the contrary, Winos
produced by this process decay, resulting in events with large missing
$E_T$ and possibly multi-leptons.  From the negative search for such
event, we may be able to discriminate the possibility of Bino-LSP.
Another possibility is to tag the flavor of quarks emitted by the
decay of gluino, which may be possible for third-generation quarks.
In the Wino-LSP case, the following process may occur:
$\tilde{g}\rightarrow\tilde{B}b\bar{b}\rightarrow\tilde{W}^+ l^+\nu
b\bar{b}$ while, for the Bino-LSP case,
$\tilde{g}\rightarrow\tilde{W}^+ b\bar{t}\rightarrow\tilde{B}l^+\nu
b\bar{t}$.  Thus, by identifying the first decay chain using $b$-jet
tagging, we may be able to confirm that the LSP is Wino.

\subsection{Testing the anomaly-mediation model}

Now we consider how well the anomaly-mediation model can be tested.
With the measurements of gaugino masses discussed in the previous
subsections, we may be able to check if the gaugino masses satisfy the
mass relation in anomaly-mediation model discussed in Section
\ref{sec:model}.

\begin{figure}
 \centerline{\epsfxsize=0.5\textwidth\epsfbox{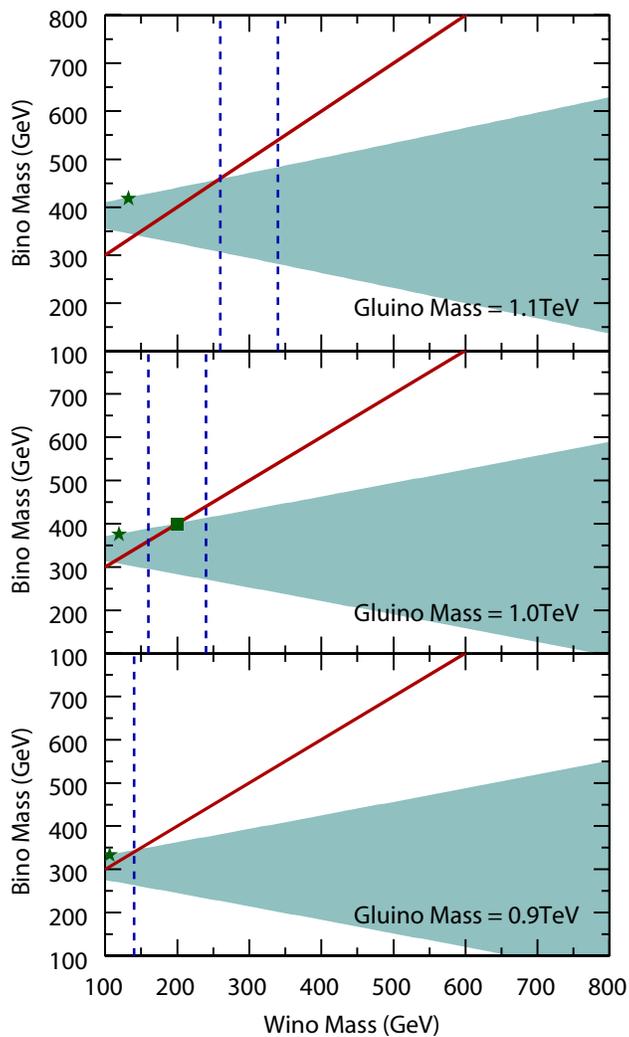}}
 \caption{Expected experimental constraints on $m_{\tilde{W}}$ vs.\
 $m_{\tilde{B}}$ plane.  The gluino mass is fixed to be $900\ {\rm
 GeV}$, $1\ {\rm TeV}$, and $1.1\ {\rm TeV}$, from below.  Solid lines
 (which are almost degenerate in the figure) are expected experimental
 upper and lower bounds on $m_{\tilde{B}}$ as functions of
 $m_{\tilde{W}}$, while the dashed lines are upper and lower bounds on
 $m_{\tilde{W}}$.  In the shaded region, gaugino masses are consistent
 with the prediction of the anomaly-mediation model.  The square in the
 middle figure shows the underlying gaugino masses used in our
 numerical analysis.  In addition, with the gluino mass being fixed, the
 Wino and Bino masses can be calculated in the case of pure anomaly
 mediation; such point is shown by the star.}  \label{fig:m2m1LHC}
\end{figure}

For this purpose, we first summarize the experimental constraints.  For
a given set of gluino masses, we plot the regions allowed by the
expected LHC constraints on the $m_{\tilde{W}}$ vs.\ $m_{\tilde{B}}$
plane.  Here, based on the discussion given in the previous section, we
assume that $m_{\tilde{B}}-m_{\tilde{W}}$ and
$m_{\tilde{g}}-m_{\tilde{W}}$ can be experimentally determined with the
errors of 1\ \% (i.e., 2 GeV) from $M_{l^+l^-}$ and 5\ \% (i.e., 40 GeV)
from $M_{l^+l^-jj}$ and $M_{jj}$, respectively.  The result is shown in
Fig.\ \ref{fig:m2m1LHC}.  (Notice that $m_{\tilde{B}}-m_{\tilde{W}}$ is
determined so accurately that the upper and lower bounds on
$m_{\tilde{B}}$ shown in the figure look almost like a single line.)  On
the same figure, we also show parameter region which is allowed in the
anomaly-mediation model.  (See the discussion at the end of Section
\ref{sec:model}.)  One of the crucial test of the anomaly mediation is
to see if those three regions overwrap for a gluino mass consistent with
experimental constraints.

The gluino mass should be constrained by using the cross section for
the process $pp\rightarrow\tilde{g}\tilde{g}$, as discussed in the
previous section.  If we can experimentally confirm that the gluino
production cross section is consistent with the predicted value for
the case of $m_{\tilde{g}}\simeq 1\ {\rm TeV}$, we can conclude that
the gaugino mass spectrum is consistent with the prediction of the
anomaly-mediation model.  

We also note that, in the present case, it is possible to exclude the
pure anomaly-mediation model (i.e., $L=0$).  In the case of pure anomaly
mediation,
the Wino and Bino masses are determined once the gluino mass is fixed.
Then, with the precise information on $m_{\tilde{B}}-m_{\tilde{W}}$
from $M_{l^+l^-}$, prediction of the pure anomaly mediation can be
excluded for the underlying parameter used in the current analysis.  In
general, accurate determination of $m_{\tilde{B}}-m_{\tilde{W}}$ will be
very useful to exclude (or confirm) the prediction of the pure
anomaly-mediation model.

\section{Summary}

In this letter, we have discussed how and how well the anomaly-mediation
model can be studied at the LHC.  As we have seen, the masses of gluino
and Wino may be constrained by using invariant-mass distribution of
dijet as well as the cross-section information.

On the contrary, the Bino mass is hardly studied in a large fraction of
parameter space since the dominant decay modes of Bino are expected to be
$\tilde{B}\rightarrow \tilde{W}^\pm W^\mp$
and $\tilde{B}\rightarrow\tilde{W}^0 h_{\rm SM}$.  Therefore, it is very
difficult to test the gaugino mass relation in the anomaly-mediation
model at the LHC.  However, in a special case where the sleptons are
much lighter than the Higgsino, the decay mode $\tilde{B}\rightarrow
\tilde{W}l^+l^-$ may acquire a large branching ratio.  In such a case,
the mass difference between Bino and Wino can be well determined by using
invariant-mass distribution of dilepton and hence the gaugino-mass
relation may be tested.

In any case, in order to confirm that Wino is the LSP, it will be
important to find the charged-Wino track.  Thus, if the excess of \ \mET
event is observed without the discovery of any sfermions, the search for
the short-lived charged-Wino track (with the decay length of $O(10\ {\rm
cm})$) is strongly suggested.

{\it Note added in proof:} 
The present procedure can be also used in models other than
anomaly-mediation model, as far as all the sfermions are heavier than
the gauginos. For example, one may test the grand-unified-theory relation among the
gaugino masses.

{\it Acknowledgement:} The authors would like to thank M. Ibe for
useful discussion at the early stage of the project.  We have used
{\tt MadGraph/MadEvent} packages \cite{MGME} for some of our numerical
calculations.

\end{document}